# Decision-Focused Learning for Future Power System Decision-Making under Uncertainty

Ran Li, *Member, IEEE*, Haipeng Zhang, Mingyang Sun, *Member, IEEE*, Can Wan, *Senior Member, IEEE*, Salvador Pineda, *Member, IEEE*, Georges Kariniotakis, *Senior Member, IEEE*, Wangkun Xu, *Student Member*, *IEEE, and* Fei Teng, *Senior Member IEEE*

*Abstract*—More accurate forecasts may not necessarily lead to better decision-making. To address this challenge, decision-focused learning (DFL) has been proposed as a new branch of machine learning that replaces traditional statistical loss with a decision loss to form an end-to-end model. Applications of DFL in power systems have been developed in recent years. For renewable-rich power systems, uncertainties propagate through sequential tasks, whereas traditional statistical-based approaches focus on minimizing statistical errors at intermediate stages but may fail to support optimal decisions at the final stage. This paper first elaborates on the mismatch between more accurate forecasts and more optimal decisions in the power system caused by statistical-based learning (SBL) and explains how DFL resolves this problem. Secondly, this paper extensively reviews DFL techniques and their applications in power systems while highlighting their pros and cons in relation to SBL. Finally, this paper identifies the challenges to adopting DFL in the energy sector and presents future research directions.

## I. INTRODUCTION

FORECAST-then-optimize is a widely-adopted framework within the field of power system operations and planning tasks, for example, day-ahead load forecasting followed by unit commitment [1], long-term wind power forecasting followed by transmission system expansion [2], and price forecasting followed by demand response [3]. Within this framework, the forecasts directly determine the optimality of downstream decision-making. For improved decisions, machine learning (ML) models have been extensively employed for the task of forecasting. Most of the existing approaches rely on statistical-based learning (SBL), where the goal is to maximize accuracy by minimizing the statistical loss, such as mean squared error (MSE) for deterministic models or pinball loss for probabilistic models. This framework builds on the assumption that lower statistical error of forecasts leads to more optimal decisions. However, recent studies have demonstrated that this assumption may not hold in the evolving landscape of power systems [4]-[6]. This is due to the fact more complex decision-making problem introduces the asymmetric, non-monotonic, and nonlinear impact of forecast error on the optimality of decision-making, which poses challenges in translating a high-quality statistical-based forecast into a high-quality decision.

Addressing these challenges necessitates a shift from SBL towards DFL, which has different names such as smart predict-then-optimize [7], integrated forecasting and optimization [8], and decision-focused forecasting [9]. This concept was first introduced by Yoshua Bengio over two decades ago [10]. Since then, it has found practical applications in various domains such as finance [11], maritime transportation [12], and computer vision [13], with promising improvement on decision-making. With the growth of renewable energy sources, decision-making under uncertainty has become critical in power systems, leading to a series of attempts at DFL in the energy sector. In order to promote the application of DFL in the energy sector, the IEEE Power and Energy Society has launched a hybrid energy forecasting and trading competition [14], which has made DFL almost an interesting industry awareness. Therefore, a review of the theory and techniques related to DFL is not only opportune but also crucial. Currently, some progress has been made in the technical review of DFL. References [15] and [8] elaborate on the classic technology of DFL. As a complement, Reference [16] clarifies the theoretical basis, benchmarks, and state-of-the-art technologies of DFL, and proposes a categorization of existing technologies. However, a systematic review, comprehensive comparison, and future directions of DFL in power systems are yet to be established. To fill this gap, we present a review of DFL application in the power system, primarily consisting of four sections to: a) elaborate on the mismatch between more accurate statistical-based forecasting and more optimal decision-making in the power system and classify them into three forms of mismatch: asymmetry, non-linearity, and non-monotonicity. b) present the fundamental principle of DFL and techniques to achieve it.

Ran Li and Haipeng Zhang are with the Key Laboratory of Control of Power Transmission and Conversion, Ministry of Education, and Shanghai Non-Carbon Energy Conversion and Utilization Institute, Shanghai Jiao Tong University, Shanghai 200240, China (e-mail: sjtu2270150@sjtu.edu.cn; rl272@sjtu.edu.cn).
Mingyang Sun is with the State Key Laboratory of Industrial Control Technology and the College of Control Science and Engineering, Zhejiang University, Hangzhou 310027, China (e-mail: mingyangsun@zju.edu.cn).
Can Wan is with the College of Electrical Engineering, Zhejiang University, Hangzhou 310027, China (e-mail: canwan@zju.edu.cn).
Salvador Pineda is with the OASYS Research Group, University of Malaga, 29071 Malaga, Spain (e-mail: spinedamorente@gmail.com).
Georges Kariniotakis is with the Center for Processes, Renewable Energies and Energy Systems (PERSEE), MINES ParisTech, PSL University, 06904 Sophia-Antipolis, France (georges.kariniotakis@mines-paristech.fr).
Wangkun Xu and Fei Teng are with the Department of Electrical and Electronic Engineering, Imperial College London, SW7 2AZ London, U.K. (wangkun.xu18@imperial.ac.uk; f.teng@imperial.ac.uk).

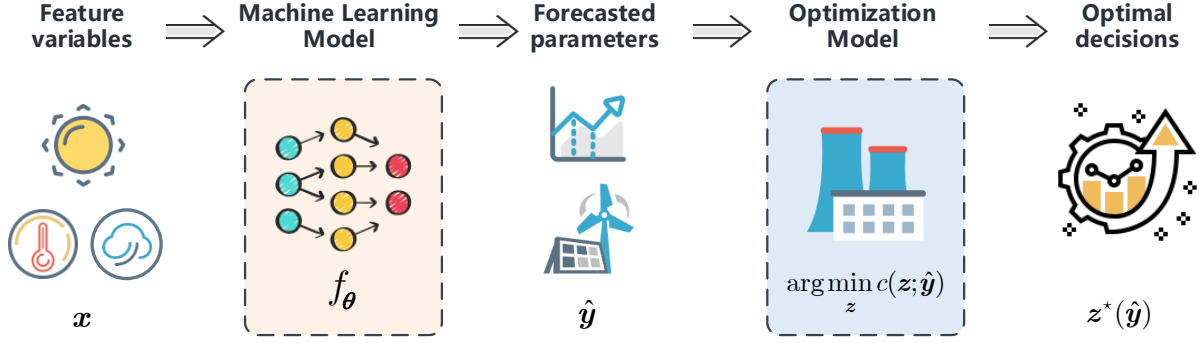

**Fig. 1.** Forecast-then-optimize framework for decision-making under uncertainty. In the forecasting stage, machine learning models are employed to forecast uncertain parameters based on features. Subsequently, in the optimization stage, optimal decisions are derived by solving the constrained optimization model utilizing the forecasted parameters.

c) extensively review various applications of DFL in the power system and identify its pros and cons in relation to traditional SBL. d) discuss the limitations of existing DFL for complex problems in power systems and characterize future developments.

## II. PROBLEM STATEMENT

The problem of decision-making under uncertainty can be solved through a forecast-then-optimize framework. The framework can be described in Fig. 1, where an ML model $f_\theta$ with trainable parameters $\theta$ is first used to forecast uncertain parameters from features $x$ (also called contextual information):

$$\hat{y} = f_\theta(x) \quad (1)$$

The forecasted parameters $\hat{y}$ are then used as input to an optimization model whose goal is to make an optimal decision $z^\star$ to minimize the objective function $c$ while satisfying a set of inequality constraints $g$ and a set of equality constraints $h$:

$$z^\star(\hat{y}) = \arg\min_z c(z; \hat{y}) \quad (2.2)$$

$$\text{s.t.} \ g(z; \hat{y}) \leq 0 \quad (2.3)$$

$$h(z; \hat{y}) = 0 \quad (2.4)$$

In the forecast-then-optimize framework, the forecasts of the ML model for uncertain parameters directly determine the optimality of the decision. Traditionally, SBL is the most common paradigm for training ML models, which separates the ML model and the subsequent optimization model so that the optimality of decision-making does not affect the training of the ML model. Specifically, SBL trains ML models by minimizing the forecast error (the deviation between the true parameters $y$ and the forecasted parameters $\hat{y}$), which can be measured by statistical loss functions such as MSE:

$$MSE(\hat{y}, y) = \|\hat{y} - y\|^2 \quad (3)$$

Despite traditional belief suggesting that lower forecast errors lead to better decisions, this may not hold in practice. Fig. 2 gives asymmetric, non-linear, and non-monotonic examples to illustrate the divergence between more accurate forecasting and more optimal decisions (see Cases 1, 2, and 3 in the Appendix for specific models). In the asymmetric example, the cost increment under a -3% load forecast error is higher than that under a 5% forecast error. A -3% error might statistically be a "better" forecast, but the resulting decision-making may prove more optimal under a 5% forecast error. This is a common scenario in power system dispatch as if the day-ahead load is under-forecasted, there may not be enough online generators leading to load shedding and consequently the expensive penalty at the value of lost load (VOLL). The example in Fig. 2(c) shows the scenario where cost increments increase nonlinearly, indicating the same forecast change may have different impacts. When the incremental forecast error increases from 2.5% to 5%, the cost is much higher compared to increasing from 0% to 2.5%. This is particularly the case for power systems with high penetration of renewables, which have a zero-marginal cost. If the forecast error can be balanced by renewables or other cheap units, the cost increment is low. Once the forecast error bounces out of the comfort zone, expensive units, which may be originally offline, have to be turned on. The last example would be counterintuitive, a perfect forecast with 0% error may not lead to the best decision. In the non-monotonic example shown in Fig. 2(d), the cost increment under a -3% forecast error is actually lower than that under a 0% forecast error. The reason is that power system dispatch is usually a two-stage decision-making process with day-ahead procurement and intraday balance. According to the regulation, a certain amount of reserve has to be purchased at the day ahead stage to ensure system reliability and stability. As a result, a perfect forecast would cost more (i.e., the unused reserve) than an under-forecast that just uses up all reserves in the intraday balance. Although this "strategic forecast" may sacrifice the reliability of power supply, the statistic tells us that a "zero-mean" forecast may not be economically optimal for multistage power system dispatch. In conclusion, transforming a good forecast into a good decision can be challenging for SBL.

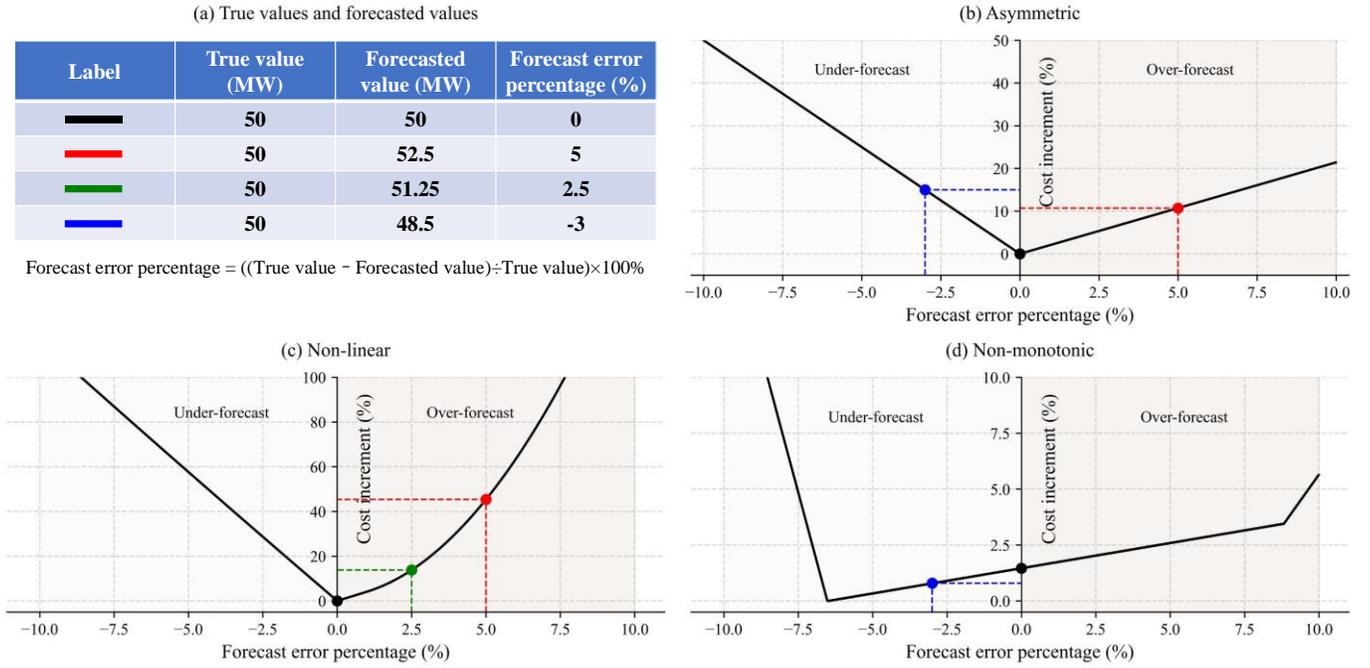

**Fig. 2.** The divergence between better forecasting and more optimal decision-making. Four different forecasts are shown in Fig. 2(a), marked in black, red, green, and blue, respectively. In Fig. 2(b), the red line is the better forecast in terms of decision benefit, since the cost increment caused by under-forecasting is higher than that caused by over-forecasting. In Fig. 2(c) The cost increment is much higher when the incremental forecast error increases from 2.5% to 5% compared with that from 0% to 2.5%. In Fig. 2(d), the relationship between negative forecasting error and cost increment is non-monotonic on the left side of the x-axis. Thus, perfect forecasting does not necessarily lead to the lowest cost increment.

## III. OVERVIEW OF DECISION-FOCUSED LEARNING

DFL is a novel learning paradigm to view sequential tasks in a holistic manner and train ML models with the ultimate goal of better decisions for the whole system. It essentially considers the decision optimality of downstream tasks for the training of the upstream model, and thus forms a closed-loop structure. This can be achieved through the utilization of a decision loss function $\mathcal{L}$:

$$\mathcal{L}(z^{\star}(\hat{y}),y) = MSE(c(z^{\star}(\hat{y});y) - c(z^{\star}(y);y)) \quad (4.1)$$

or

$$\mathcal{L}(z^{\star}(\hat{y}),y) = MSE(z^{\star}(\hat{y}) - z^{\star}(y)) \quad (4.2)$$

The decision loss measures the deviation between the optimal decision objective $c(z^{\star}(\hat{y});y)$ under the forecasted parameters $\hat{y}$ and the optimal decision objective $c(z^{\star}(y);y)$ under the true parameters $y$. Decision losses can be in different forms, such as regret [9], expected objective [17] and SPO loss [7], etc. Through these decision loss functions, the ML and optimization models are integrated into a whole process, allowing decision optimality to explicitly affect the ML model training. DFL can be broadly divided into indirect approaches and direct approaches according to whether it produces explicit forecasts, as shown in Fig. 3.

### A. Indirect Approach

The training process of the ML model in the indirect method includes the forward propagation process and the backward propagation process, as shown in Fig. 3(a). During the forward propagation process, the ML model obtains the forecasted parameters from the features $x$, which is then used as input by the optimization model to obtain the forecasted optimal decision $z^{\star}(\hat{y})$. During the backward propagation process, the forecasted optimal decision and the real optimal decision $z^{\star}(y)$ constitute the decision loss $\mathcal{L}(z^{\star}(\hat{y}),y)$, and then the gradient of the decision loss is calculated and the parameters $\theta$ in the ML model are updated through gradient descent. The gradient of the decision loss can be split into three parts $\partial \mathcal{L}/\partial z^{\star}(\hat{y})$ (①), $\partial z^{\star}(\hat{y})/\partial \hat{y}$ (②), and $\partial \hat{y}/\partial \theta$ (③) through the chain rule, of which ① and ③ are relatively easy to calculate because they are differentiable. The difficulty in calculating decision loss is due to the fact that ② may be discontinuous, non-differentiable, or even zero-valued depending on different types of optimization models:

**Convex optimization**: convex optimization problems represent a specialized type within mathematical optimization. These problems are distinguished by the unique property that both the objective function and the constraints exhibit convexity [18]. This property ensures that the locally optimal solution is actually globally optimal. Furthermore, many solution methods to convex optimization problems can be proven to converge towards local optima. However, within the scope of DFL, computing the gradient of a convex optimization model remains a challenge. Consider linear programming (LP), a subfield of convex optimization, where optimal solutions consistently lie at the vertices of the feasible domain. When translating continuous parameters into discrete solutions, this process results in a piecewise constant mapping. Therefore, the derivatives of the LP are calculated to zero at these vertices and undefined elsewhere, which is not conducive to gradient-descent-based training [15].

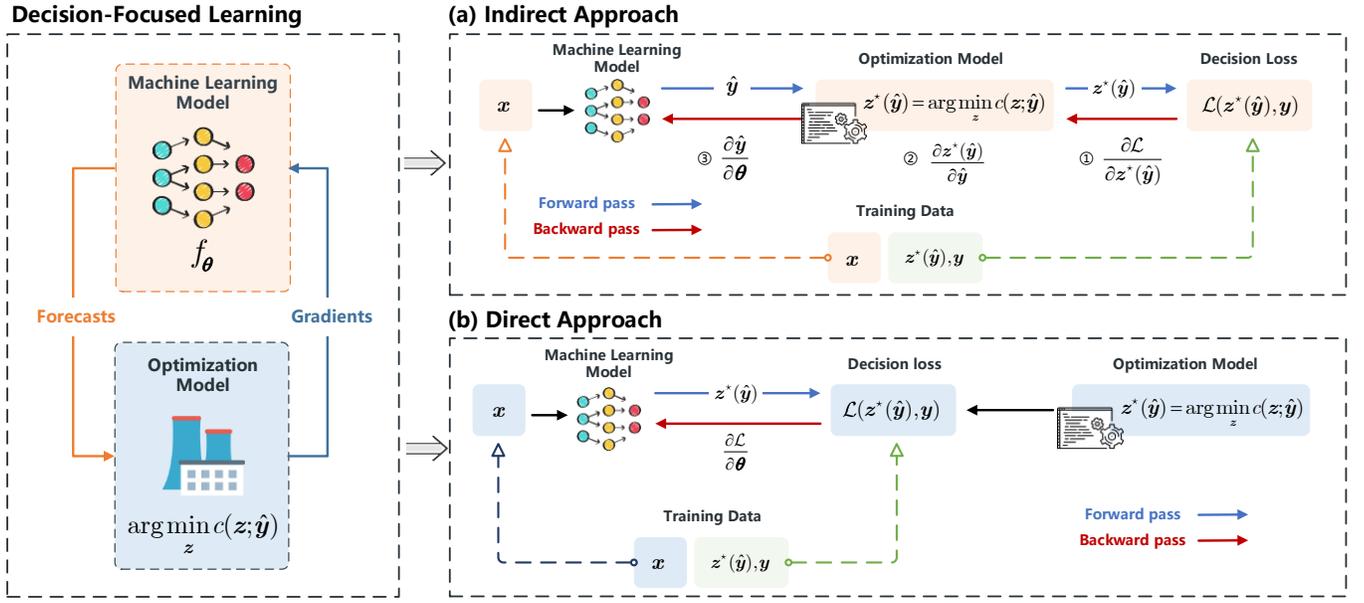

**Fig. 3.** Direct and indirect approaches to achieve decision-focused learning. Decision-focused learning is a learning paradigm that trains ML models in a forecast-then-optimize framework with decision optimality as the ultimate goal. The indirect approach follows the process of forecasting first and then optimizing, explicitly generating forecasts, while the direct approach simplifies this process and does not explicitly generate forecasts.

**Integer linear programming/mixed integer linear programming**: integer linear programming (ILP) and mixed integer linear programming (MILP) are extensions of LP in which the decision variables are fully or partially constrained to integer values. Compared to LP, incorporating ILP/MILP into the DFL context presents heightened challenges. On one hand, their inherent NP-hard property may impede the precise computation of the optimal decision solution $z^*(y)$, rendering the acquisition of accurate gradients a formidable task. On the other hand, the presence of discontinuities within their feasible regions exacerbates the inherent discontinuity and non-differentiability of the decision loss function [9].

**Generic optimization problems**: Generic optimization problems refer to all types of optimization problems. Existing works have developed one-size-fits-all DFL methods for general optimization problems, such as Wang et al. [19], and Li et al. [20], although the properties of different optimization problems lead to different challenges in obtaining gradients. However, these methods have only been validated in applications with linear objectives, and their performance in more complex applications remains to be validated.

Another challenge in computing ② lies in the substantial computational cost it imposes. The $z^*(\hat{y})$ in ② means that one optimization problem needs to be solved for each training instance at each epoch. Even when dealing with convex optimization problems, the computational complexity is typically orders of magnitude greater than that of traditional neural network layers. When addressing ILP or MILP problems, their inherent NP-hard property further escalates the computational costs [21].

Many methods have been developed to address the above two key challenges, which can be roughly divided into four categories according to the taxonomy proposed by Mandi et al. [16]:

**Differentiation method**: In this category, the goal is to calculate exact gradients by differentiating the optimal conditions of an optimization problem where the derivatives are well-defined and non-zero. For example, convex quadratic programming can utilize implicit differentiation of the Karush-Kuhn-Tucker (KKT) conditions to compute the derivatives of the solution with respect to parameters.

**Smooth method**: In this category, the goal is to generate an approximate gradient by employing a smooth approximation of the differentiation of the optimization problem. For instance, while the differentiation of an LP problem may be either zero or undefined, useful gradients can be obtained to guide ML model training through the application of smoothing methods, such as the introduction of regularization terms.

**Perturbation method**: In this category, the goal is to achieve smooth approximation by introducing random perturbations to discrete solutions. From this perspective, $z^*(\hat{y})$ can be viewed as a random variable, conditionally dependent on $y$. The derivative of the expectation of $z^*(\hat{y})$ with respect to the parameters exists and is non-zero, so the gradient problem can be easily solved.

**Surrogate method**: In this category, the goal is to propose a convex surrogate loss function for decision loss functions (such as regret). For example, the discontinuous and non-differentiable regret decision loss can be replaced by a parameter-defined convex function (such as a linear piecewise

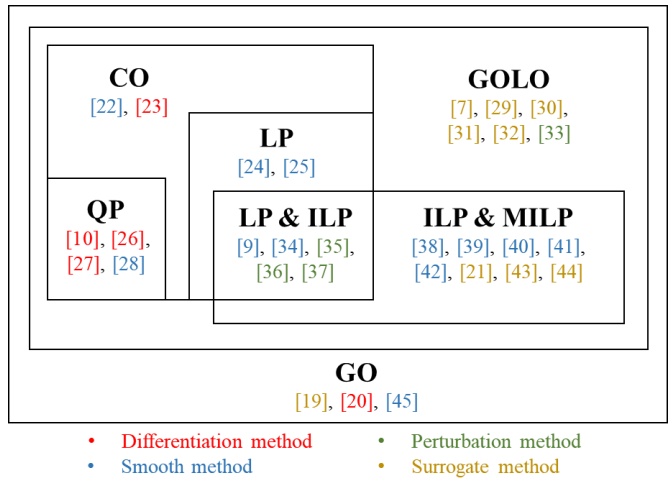

**Fig. 4.** A taxonomy of existing indirect approaches. Red represents the differentiation method, green represents the perturbation method, blue represents the smooth method, and orange represents the surrogate method. Optimization problems are divided into convex optimization (CO), quadratic programming (QP), linear programming (LP), integer linear programming (ILP) and mixed integer linear programming (MILP), generic optimization (GO), and generic optimization with linear objectives (GOLIN).

function), thereby providing useful gradients to guide ML model training.

An alternative approach to traditional forward and backward propagation is to reformulate optimization problems (4.1) and (4.2) as bilevel optimization problems. Indeed, the function $z^\star(\hat{y})$ is defined through the optimization problem (2) and serves as the lower-level problem within a bilevel model framework, as outlined in [6]. While solving bilevel problems can be computationally demanding, the technical literature offers efficient methods tailored for specific instances of lower-level optimization problems. In [6], the authors propose a non-convex single-level reformulation for a quadratic lower-level problem, along with a linear mapping between the context and the uncertain parameters. This single-level problem can be addressed using a regularization iterative algorithm or further reconfigured as a mixed-integer program, incorporating additional binary variables and large enough constants.

For the convenience of readers, Fig. 4 illustrates the types of optimization problems to which existing indirect methods are applicable, along with their respective method classifications.

*B. Direct Approach*

Compared with indirect approaches, the characteristic of direct approaches is that there is no need to explicitly generate forecasted parameters [46]. In other words, in the forecast-then-optimize framework, the ML model directly generates forecasted optimal decisions instead of forecasted parameters from features. Training the ML model through the direct method is also divided into forward propagation and backward propagation processes, as shown in Fig. 3(b). In the forward propagation process, the ML model directly generates the forecasted optimal decision $z^\star(\hat{y})$ from the feature $x$. In the backward propagation process, the parameters in the ML model are updated with the goal of minimizing the decision loss $\mathcal{L}(z^\star(\hat{y}), y)$.

Rudin and Vahn [46] proposed and investigated the application of this approach to the newsvendor problem. Unlike other formulations, the newsvendor problem does not involve any constraints. Therefore, the optimal decisions obtained are always valid and reliable. Carriere and Kariniotakis [47] utilized this approach in addressing the energy trading problem. In this work, they established a mapping relationship between photovoltaic power and day-ahead market bids, employing neural networks. Lu et al. [48] applied this approach to economic dispatch and underscored its effectiveness in reducing additional costs. However, it is important to note that the indirect approach may not guarantee feasible solutions for out-of-sample observations in various practical applications where optimal decisions need to comply with a specific set of constraints.

Methods to mitigate the feasibility predicament have been proposed and fall into two distinct categories. The first category is the post-processing method, which employs verification and correction techniques to guide the forecasted decisions into the feasible region delineated by the optimization model [49], [50]. The second group is the regularization method, which entails transforming the optimization model's constraints into regularization components, which are then integrated into the loss function of the ML model [51], [52]. To ensure the forecasted decisions remain within the feasible domain, substantial weights are assigned to these regularization terms. However, the above methods still cannot perfectly solve the feasibility problem. While the post-processing methods guarantee the feasibility of optimization results, they may compromise the optimality of results. The regularization methods make commendable efforts to keep optimization results within the feasible region but do not entirely eliminate the possibility of them falling outside.

In order to ensure that the forecasted optimal decisions always fall within the feasible region, Bertsimas and Kallus [17] proposed to train the ML model through conditional stochastic programming based on the sample average approximation (a widely recognized data-driven solution strategy in stochastic programming). This method is compatible with a variety of ML algorithms, such as k-nearest neighbor regression, local linear regression, random forest, etc. Bertsimas et al [53]. demonstrated that decision trees are an optimal choice for implementing the prescriptive analytics method due to their interpretability, scalability, and generalizability. However, the prescriptive analytics method can be susceptible to overfitting issues when confronted with noisy data, which raises uncertainties about its out-of-sample performance guarantees. To tackle this challenge, Bertsimas et al. [54] proposed a robust optimization and statistical bootstrap-based technique to prevent overfitting. In a similar vein, Srivastava et al. [55] provided performance guarantees for out-of-sample scenarios employing techniques derived from moderate deviance theory. Recently, Stratigakos et al.

TABLE I
REVIEW OF DFL APPLICATIONS IN POWER SYSTEMS

| Reference | Method | Forecast | Optimize | Data |
|---|---|---|---|---|
| Stratigakos et al. (2021) [57] | Indirect | Imbalance penalty (1 decision tree model) | Energy trading (LP) | German electricity market |
| Chen et al. (2022) [58] | Indirect | Renewable power (4-5 linear regression models) | Unit commitment (MILP) | Belgian Independent System Operator (ISO) [74] |
| Zhang et al. (2022) [59] | Indirect | Load (1 linear regression or neural network model) | Energy trading (QP+QP) | Global energy forecasting competition 2012 [75] |
| Wang et al. (2017) [60] | Indirect | Load (1 neural network model) | Unit commitment (MILP) | New York City ISO [76] |
| Sang et al. (2021) [61] | Indirect | Electricity price (1 neural network model) | Storage arbitrage (MILP) | Pennsylvania-New Jersey-Maryland (PJM) interconnection [77] |
| Garcia et al. (2022) [62] | Indirect | load and reserve (1 linear regression model) | Energy trading (LP+LP) | Random process generation |
| Zhang et al. (2023) [63] | Indirect | Inertia (1 linear regression model) | Inertia management (QP+LP) | Great Britain power system [78] |
| Han et al. (2021) [64] | Indirect | Load (1 neural network model) | Economic dispatch (QP) | PJM interconnection |
| Morales et al. (2022) [65] | Indirect | Load (1 linear regression model) | Energy trading (LP+LP) | ENTSO-e Transparency Platform [79] |
| Xu et al. (2023)[66] | Indirect | Load (1 neural network model) | Economic Dispatch (LP) and redispatch (LP) | Texas Backbone Power System [80] |
| Muñoz et al. (2020) [67] | Indirect | Wind power (1 linear regression model) | Energy trading (LP) | Danish TSO and ENTSO-e Transparency Platform |
| Chen et al. (2022) [68] | Indirect | Renewable power and reserve (5 linear regression models) | Unit commitment and economic dispatch (MILP+MILP) | Belgian ISO |
| Zhao et al. (2022) [69] | Indirect | Wind generation interval (1 extreme learning machine model) | Optimal wind power offering (LP) | Glens of Foudland Wind Farm [81] |
| Zhao et al. (2021) [70] | Indirect | Wind generation interval (1 extreme learning machine model) | Reserve procurement (LP) | Irish transmission system operator [82] |
| Carriere et al. (2019) [47] | Indirect/Direct | Renewable power (1 extreme learning machine model) | Energy trading (LP) | European center for medium-range weather forecasts |
| Lu et al. (2022) [48] | Indirect/Direct | Load (1 long short-term memory model) | Economic dispatch (LP) | German power system [83] |
| Stratigakos et al. (2022) [71] | Direct | Renewable power (1 decision tree model) | Energy trading (LP) | ENTSO-e Transparency Platform |
| Wahdany et al. (2023) [72] | Indirect | Wind generation (1 neural network model) | Energy trading (LP) | Germany and Netherlands weather data [84] |
| Zhang et al. (2023) [73] | Indirect | Wind generation (1 extreme learning machine model) | Virtual power plant operation (LP) | Global energy forecasting competition 2012 |

[56] applied the prescriptive analytics method to the optimal power flow tasks in power systems to improve computational efficiency while ensuring the feasibility of the solution.

IV. APPLICATION OF DECISION-FOCUSED LEARNING IN POWER SYSTEMS

The literature that applies DFL to power system operational decision-making tasks is summarized in Table 1. The table provides a detailed summary of each literature, including a) objectives of the forecasting tasks and ML models employed for these forecasts. b) Decision-making tasks and types of optimization models. c) data sources of case study.

The focus of forecasting tasks in the DFL literature has been renewable output and load demand. Various uncertainty parameters stemming from the two, including imbalance penalties, reserves, and inertia, also attracted a certain degree of attention. From ML methods perspective, linear regression and neural networks emerge as the most commonly employed ML models for the forecasting task, and the ML task is usually single-stage as a single category of uncertain parameters is considered by default in decision-making tasks. Decision-making tasks focus on energy dispatch (e.g., unit commitment and economic dispatch) and electricity market clearing (e.g., energy trading and reserve procurement), where most of optimization models are one-stage, while some of the literature considers corrective actions after the realization of the uncertainty parameter, constituting a two-stage optimization. At the methodological level, the indirect method is more widely adopted compared to the direct method. This may be due to the fact that the direct method hides the forecast of uncertain parameters, resulting in potential disputes in tasks involving multiple stakeholders. The comparative findings of DFL and SBL in each work are summarized in Table 2.

TABLE II
FINDINGS OF DFL APPLICATIONS IN POWER SYSTEMS

| Reference | Findings | Reference | Findings |
|---|---|---|---|
| Stratigakos et al. (2021) [57] | For the case of a virtual power plant located in France, the total revenue from energy trading increased by an average of 0.58%, and the operating risk decreased by an average of 14.4%. | Muñoz et al. (2020) [67] | In the Danish DK1 bidding zone, the accuracy of wind power forecasts improved by 8.53%, and the balance costs incurred by wind power producers decreased by 2.13%. |
| Chen et al. (2022) [58] | In an IEEE RTS 24-bus system and an ISO-scale 5655-bus system, the average improvement in the daily day-ahead cost over an entire year could be as high as 1.96%. | Chen et al. (2022) [68] | In an IEEE 118-bus system, the daily market economy improved by an average of 0.82%, although the forecast accuracy was slightly worse. |
| Zhang et al. (2022) [59] | In an IEEE 33-bus system, the economic benefits increased by an average of 13.74%. | Zhao et al. (2022) [69] | In China's time-of-use electricity price scenario, the operating costs of wind power are decreased by at least 9%, and the low coverage deviation is kept within ±2.1% |
| Wang et al. (2017) [60] | In a 3-area 72-bus system, the forecast error increased by 1.9% on average, but the day-ahead cost decreased by 11.33% on average. | Zhao et al. (2021) [70] | In Irish systems, reserve costs are decreased by at least 2% compared to other traditional interval forecasting methods. |
| Sang et al. (2021) [61] | Regret decreased by 90.91% and economic benefits increased by 46.8%. | Carriere et al. (2019) [47] | Compared with the indirect method, the training time of the ML model and decision time are significantly reduced, but the gain in economic benefits will be slightly reduced. |
| Garcia et al. (2022) [62] | In a 3000-bus system and a 300-bus system, the economic benefits are increased by 13% and 11.4% respectively. | Lu et al. (2022) [48] | In an IEEE 39-bus system, the additional cost of inaccurate forecasts was reduced by 5.49%, while the training speed was increased by 500 times compared with the indirect method. |
| Zhang et al. (2023) [63] | In the GB 2030 system, the average system cost decreased by 1.81%, and the average system risk decreased by 7.89%. | Stratigakos et al. (2022) [71] | In the French electricity market, the total profit for electricity market trading increased by 3.82% and 0.62% on average under the single- and dual-price balancing mechanism, respectively. |
| Han et al. (2021) [64] | In a 3-bus system and an IEEE 118-bus system, the system costs are decreased by 0.24% and 0.27% respectively, but the forecast errors increase by 0.36% and 0.16% respectively. | Wahdany et al. (2023) [72] | In an IEEE 6-bus system, forecast errors of system costs are reduced by up to 10% with high wind capacity. Wind curtailment was reduced by more than 20% for individual cases but increased overall. |
| Morales et al. (2022) [65] | In the 28-node European electricity market model, the economic benefits achieved are much higher than 2%. | Zhang et al. (2023) [73] | Although DFL has the largest Winkler score and negative average coverage deviation, indicating poor forecast accuracy, it has the lowest average system cost, indicating better decisions. |
| Xu et al. (2023) [66] | Ignoring the unpredictable uncertainties in the optimizations may cause high costs in real-time. Adversarial training improves the robustness of the DFL. | | |

Based on findings in the existing literature, this paper identifies six key features to evaluate DFL and SBL and provides a detailed comparison. SBL can be broadly divided into two groups. The first group is the deterministic approach, which generates point forecasts. This method focuses on providing a single value that represents the most likely outcome. The second group is the probabilistic approach, which generates probability distributions as outputs. This method emphasizes capturing the inherent uncertainty in the forecast, offering a range of possible outcomes along with their associated probabilities. The potential is marked from one plus sign "+" to three plus signs "+++" representing from low to high.

**Offline training cost**: Offline training cost refers to the computational cost required to train a parameter forecasting or decision forecasting model. In this regard, DFL incurs high computational costs due to the incorporation of optimization models, which are usually large in size and contain integer variables as well as nonlinear constraints [7]. By contrast, SBL is relatively inexpensive in terms of computational cost, owing to the advanced ML technology currently available.

**Online decision cost**: Online decision cost refers to the computational cost required to make a decision. The indirect method and the deterministic method exhibit similar computational costs since both rely on deterministic optimizations to derive subsequent decisions. The direct method is considerably less costly because the decision is obtained through a specific function [47]. Since the probabilistic method generates not a value but a distribution, its downstream optimization model will be based on stochastic programming or robust optimization, resulting in the highest online computational cost [1].

TABLE III
EVALUATE THE POTENTIAL OF THE FOUR METHODS IN SIX FEATURES

| Features / Approaches | Offline training cost | Online decision cost | Decision optimality | Transparency | Customizability | Portability |
|---|---|---|---|---|---|---|
| Deterministic approach | + | ++ | + | +++ | + | +++ |
| Probabilistic approach | ++ | +++ | +++ | ++ | +++ | +++ |
| Indirect approach | +++ | ++ | +++ | +++ | ++ | + |
| Direct approach | +++ | + | +++ | + | ++ | + |

**Decision optimality**: Decision optimality refers to the positive outcome or advantage gained by making a particular decision. DFL and the probabilistic method are superior to the deterministic method in terms of decision benefit, as the latter is incapable of considering the impact of forecast errors on the decision objective [68]. A typical paradigm of probabilistic methods is to employ probabilistic ML models to generate possible scenarios and optimize overall expectations through stochastic programming, which can also consider the impact of forecast errors on decision optimality, thereby achieving better decisions.

**Transparency**: Transparency refers to observability throughout the decision-making process. The indirect and deterministic methods exhibit strong transparency since the entire process, from forecasting the parameter value to making a decision based on the forecast, is fully transparent. The probabilistic method involves defining and selecting scenarios, which reduces transparency to a certain extent [1]. In tasks involving multiple stakeholders, such as electricity markets, it may not be conducive to reaching a consensus among the stakeholders, as issues concerning the definition of scenarios, number of scenarios, and probability of scenarios will be highly contentious. The direct method employs a black-box ML model to directly forecast the decision, thus hiding the entire process from forecasting to decision-making and significantly affecting transparency [53].

**Customizability**: Customizability refers to the ability of a method to be customized to meet the risk preferences of decision-makers. The probabilistic method can achieve customizability by introducing risk measures or robust optimization. While the direct and indirect methods are well-suited for risk-neutral individuals, their lack of customizability has been a notable limitation. However, recent studies have explored the integration of risk measures into these methods to achieve customizability. Consequently, both methods now have the potential to cater to the preferences of decision-makers with varying levels of risk preference [63]. It is worth noting, however, that the deterministic method remains non-customizable.

**Portability**: Portability refers to the ability of parameter forecasting or decision forecasting tools to be easily transferred from one decision task to another without major adjustments. The portability of DFL is limited, as both methods take into account downstream decision-making tasks when training forecasting models. In contrast, SBL does not consider the downstream decision-making task, thus exhibiting superior portability.

V. CHALLENGES AND RESEARCH DIRECTIONS

Most of the forecast-then-optimize frameworks in the literature consider a single deterministic ML model paired with a single deterministic optimization model, which is defined as the default setting. However, forecasting and decision-making tasks in power systems may be more complex, and technologies under this default setting may not be directly applicable to these tasks. We illustrate this point from three aspects: forecasting tasks, decision-making tasks, and other aspects.

*A. Forecasting Task*

The challenges in forecasting tasks can be summarized in Fig. 5.

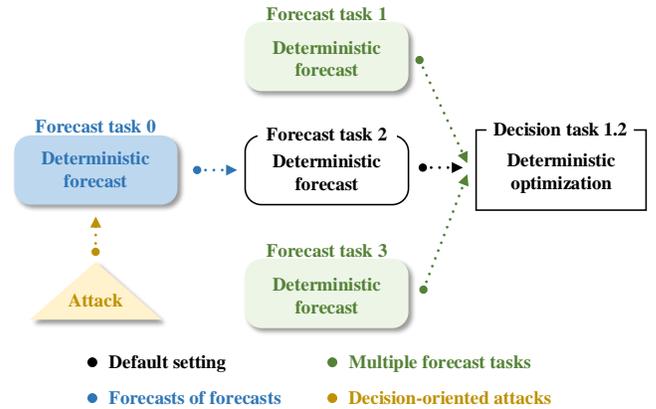

**Fig. 5.** Challenges in forecasting tasks.

**Multiple forecast tasks**: Decision-making processes within the domain of power systems often rely on predictions generated by multiple ML models. For instance, if one participates at multiple markets and manages storage in the very short term, it concerns both forecasting (multiple forecasts for different time frames) and also decision-making. In this setting, DFL needs to simultaneously train multiple ML models, resulting in a significant increase in the number of parameters to be updated, making overfitting highly prone to occur. Therefore, techniques to prevent overfitting become crucial. However, the existing literature offers limited insights into this specific problem, and technologies commonly employed are still those frequently used in SBL, such as L1 or L2 regularization [57]. Although these techniques have demonstrated effectiveness in the SBL domain, their performance in the DFL domain remains a subject of uncertainty and requires further investigation. Another challenge to forecasting is to combine multiple forecast products that are not homogenous like probabilistic forecasts, ramp forecasts, risk indices et in decision making. So far for example many transmission system operators have developed renewable energy sources ramp forecasting but it is like a dead end since they cannot integrate it into their decision-making efficiently.

**Forecasts of forecasts**: In power systems, the utilization of forecasts in one forecast task as input features for another forecast task is a common setting. For instance, meteorological information derived from numerical weather prediction serves as a crucial input feature for short-term wind power forecasting models, and load and weather forecasts are employed as input features for electricity price forecasting models. However, when employing DFL in this setting, a significant challenge arises, as the forecasts used as input features may not always be generated through ML models. For instance, numerical weather prediction is inherently a system of equations, making it infeasible to update it based on the gradient of decision error. There is currently a lack of literature addressing this specific problem.

**Decision-focused attacks**: ML-based forecast models within power systems exhibit a high degree of dependence on external forecast information, typically sourced from other modules and communicated through application programming interfaces. This dependence on external data renders ML-based forecast models susceptible to cyberattacks. Malicious actors can exploit this vulnerability by crafting inputs to ML models in a manner that manipulates their outputs, constituting evasion attacks. Additionally, they can generate adversarial training data that detrimentally affects the performance of trained models, constituting poison attacks. Given the escalating threat of cyberattacks, it becomes imperative to comprehensively investigate the vulnerabilities and ramifications associated with ML-based DFL before deploying them in practical applications. However, research on this topic is still in its nascent stages. Chen et al. [85] examined the vulnerability and impact of ML models in the context of decision-focused data integrity attacks. Their findings suggest that decision-focused attacks can have a more adverse impact on power system operation and are more concealed than common adversarial random attacks.

*B. Decision-making Task*

The challenges in decision-making tasks can be summarized in Fig. 6.

**Multi-stage decision-making tasks**: In power system decision-making, many tasks are executed in a sequential manner, such as starting with day-ahead unit commitment and followed by real-time economic dispatch. This setting entails that the solution of the previous decision-making task impacts the subsequent one. As a result, decision-makers may focus on the overall decision utility across all decision-making tasks. Under this setting, the forecast error will be propagated along the stages, resulting in a more complicated relationship between forecast error and decision optimality. In some cases, this can even result in a mismatch between the perfect forecast and the globally optimal decision solution, as depicted in Fig. 2(d). On the other hand, multi-stage decision-making may include multiple types of optimization problems, making it more challenging to calculate the gradient of the decision loss. To solve these problems, solutions have been proposed to integrate the ML model and multi-stage decision-making tasks into a multi-layer mathematical programming framework [68]. Mathematical programming solvers are then employed to solve the ML model parameters. This solution is well-suited to multi-stage problems and avoids the need to calculate gradients. However, this solution is computationally intensive and only compatible with linear ML models. Further research to enable more complex ones such as neural networks is still needed.

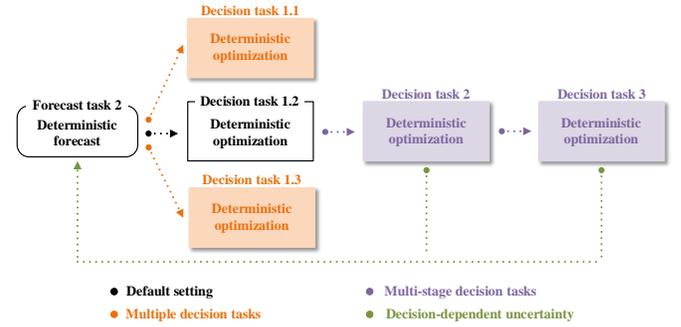

**Fig. 6.** Challenges in decision-making tasks.

**Decision-dependent uncertain parameters**: In the default setting, the decision solution has little impact on future uncertainties. Therefore, the uncertain parameters involved are called decision-independent uncertainty parameters. However, the solutions to many decision-making tasks in power systems will have an impact on the uncertain parameters involved, which are called decision-dependent uncertain parameters [86]. For example, deferrable loads consisting of electric vehicles, washing machines, dishwashers, etc. can be scheduled in a demand response program. This scheduling action shifts their power consumption to future time intervals, subsequently altering the probability distribution or the upper and lower bounds of future loads. Therefore, in a multi-stage decision-making setting, decision-dependent uncertainty may complicate the relationship between forecast error and decision optimality. There is currently very little research on this topic. Only Liu et al. [87] considered the decision-dependent uncertainty within the framework of deterministic forecast followed by stochastic programming, but there is a lack of cases in power systems for verification.

**Multiple decision-making tasks**: A forecast task may have impacts on several decision-making tasks within the power system. For instance, load forecasting tasks play a crucial role in various decision-making tasks, including generation scheduling, power market clearing, demand response, and energy storage system management. In this setting, it is an extremely time-consuming and labor-intensive endeavor to tailor a dedicated ML-based forecasting model for each decision-making task using DFL. To address this challenge, Tang et al. [88] have proposed an innovative method leveraging multi-task deep learning. This approach facilitates the sharing of information across different tasks, thereby avoiding the need to train distinct models for each individual decision-making task. However, this method has yet to be empirically validated within the context of power systems.

*C. Other Aspects*

**Uncertain parameters in constraints**: Most research in the field of DFL focuses on forecasting uncertain parameters within the objective function, assuming that parameters in the constraints are precisely known. However, in the context of power systems, unknown parameters can exist not only in the objective function but also within the constraints. For instance, consider the power balance constraint in the unit commitment problem, where the equality between the sum of renewable energy output and unit output and the load demand must hold. Within this constraint, load demand and renewable energy output are typically treated as uncertain parameters. When it comes to forecasting these parameters within the constraints and utilizing them to obtain an optimal decision, a crucial concern arises: the decision made may not be feasible with respect to the true parameters. The existing approach to address this issue involves penalization when the forecasted optimal decision becomes infeasible for the true parameters. Garcia et al. [62], Li et al. [63], and Zhang et al. [59] proposed the use of corrective actions to penalize unfeasible decisions, while Hu et al. introduced the concept of post-event regret to promote the transformation of infeasible solutions into feasible ones by minimizing post-event regret.

**Probabilistic forecasting and uncertainty optimization**: The settings of deterministic forecasting and then deterministic optimization may not always apply. Forecasting tasks in power systems may be of probabilistic in nature, and related decision-making tasks may involve uncertainty optimization rather than deterministic optimization. In fact, there is an ultimate-goal mismatch between probabilistic forecasting and uncertainty optimization, such that merely improving the statistical error of prediction intervals with respect to calibration and sharpness does not always help improve the decision optimality of a specific decision problem. To bridge this gap, Zhao et al. [69] established a novel cost-oriented ML framework that combines non-parametric renewable power prediction interval construction and decision-making. The framework is formulated as a two-level programming model that minimizes the operational cost of the decision-making task by adaptively adjusting the prediction intervals generated by quantile regression. Zhang et al. [73] focus on the integration of prediction intervals and robust optimization. Sen et al. [89] provided a paradigm called learning-enabled optimization to integrate SBL and stochastic programming. They proposed several new concepts, such as statistical optimality, a convex extension of stochastic decomposition, hypothesis tests for model-fidelity, generalization error of stochastic programming, etc., to provide support for the modelling, solving, and verification of this framework. Donti et al. [27] demonstrated how to convert the setting of deterministic forecasting and then stochastic optimization into the default setting and use the technology in DFL to solve it. Uncertainty optimization will increase the computational cost when calculating the decision loss, and data-driven methods can be considered to alleviate this problem. For example, use deep neural networks to solve stochastic and chance constraints[90], [91].

**Risk preference**: In the default setting, decision-making tasks tend to be risk-neutral. However, decision-makers in the energy sector tend to adopt a risk-averse approach to ensure the security, stability, and reliability of power systems. Consequently, the integration of DFL with consideration for risk preferences becomes more suitable for decision-making tasks in power systems. Tulabandhula et al. [92] introduced a method wherein the decision objective is transformed into a regularization term within the objective function of the learning algorithm. This approach allows for an optimistic or pessimistic perspective regarding potential costs, depending on the values of the regularization parameters. Furthermore, Zhang et al. [63] extended the concept of DFL by integrating it with risk measures. Through the manipulation of risk parameters, they aimed to align DFL with either risk-neutral stochastic programming or risk-averse robust optimization, thus offering greater flexibility for risk management of decision-making in power systems.

**More forecast-then-optimize scenarios**: As can be seen from Table 1, DFL application scenarios are mostly concentrated in power system energy dispatch. However, this technology can actually be applied to more scenarios, such as distribution network active and reactive power optimization, optimal power flow, optimized operation of energy storage systems, power market operation and pricing, power system planning and expansion, etc. For these scenarios, existing work has developed many useful tools, which can be well combined with DFL. For example, the semi-definite programming, quadratic programming, and second-order cone programming formulas of the optimal power flow model for distribution and transmission systems can be used to easily calculate the differential of the decision loss and then update the forecasting model[93]-[95].

**More than forecast-then-optimize**: Beyond the conventional task of forecast-then-optimize, there exists several two-stage tasks within the domain of power systems, such as cluster-then-forecast and cluster-then-optimize. It is worth noting that DFL has shown potential for application on these tasks, and several studies in the literature have already delved into these extensions. Zhang et al. [96] introduced an integrated approach encompassing clustering and prediction. Their proposed method achieved a significant performance enhancement of 52.20% when compared to the conventional bottom-up approach. Furthermore, Sun et al. investigated the integration of load scenario selection with transmission network expansion planning [97], while also exploring the integration of representative day selection and investment decisions within the power systems [98]. Zhang et al. [99] contributed by proposing a cost-oriented clustering technique and successfully applying it to pricing and power consumption scheduling problems. Additionally, Wogrin [100] explored the relationship between time series aggregation, a method used to reduce input data dimensionality and model size, and its impact on economic dispatch. Zhou et al. [101] utilized the DFL framework to evaluate the value of data in an end-to-end way. Xu et al. [66] used the DFL framework to unlearn part of the training data from the trained ML model with the operation cost considered.

**The trade-off between the accuracy of solving optimizations during training and the training speed**: This can be a better optimization algorithm or some distributed algorithm can be implemented. We can also sacrifice some of the accuracy to achieve faster convergence for large systems. For example, [48] was developed on the basis of [27], [64] to reduce the training burden of differentiable layers.

**Missing data**: Another challenge is to be able to have a resilient approach to missing corrupted data for forecasting and optimization. The classical approach of decision making assumes a given level of data quality but what happened when this is not granted. The ML-based DFL gives the possibility to have more flexible decision-making schemes as a function of the available data during operation. Here, Akylas et al. [102] discussed resilient forecasting based on robust optimization and the extension of the prescriptive approach for the case of missing data.

## APPENDIX

In power system decision-making tasks, the forecasts for uncertain parameters are crucial for decision optimality. The most common method is to use the expected value, as it is considered to be the best approximation of the true value. The underlying assumption is that forecasts that are closer to the true values will result in more optimal decisions. However, in practice, this assumption does not always hold, as demonstrated by the following three examples:

**Example 1**: Here we consider a single bus system with three generators (G1, G2, and G3) serving an uncertain load. Let $P = [P_1, P_2, P_3]$, $\bar{P} = [30, 30, 20]$ MW, and $C = [80, 90, 100]$ \$/MW denote the power generation, the generation limit, and the cost coefficient of the generator, respectively. Given the forecast of the uncertain load $\hat{L}$, the lowest-cost day-ahead energy schedule can be formulated as:

$$\min_{P_i} \sum_{i=1}^{3} C_i P_i \tag{1a}$$

$$\text{s.t.} \ 0 \leq P_i \leq \bar{P}_i, \quad i = 1, 2, 3, \tag{1b}$$

$$\sum_{i=1}^{3} P_i = \hat{L}, \tag{1c}$$

In the real-time stage, unbalanced loads are covered by load shedding:

$$\min_{L_{sh}} C_{VoLL} L_{sh} \tag{2a}$$

$$\text{s.t.} \ L_{sh} \geq L - \hat{L}, \tag{2b}$$

where $L$ represents the true load, $L_{sh}$ represents the load-shedding decision, and $C_{VoLL}=300$ \$/MW represents the value of the lost load. The forecast error can be controlled through sampling so that the impact of forecast error on the total cost increment is obtained, as shown in Fig. 2(b).

**Example 2**: We consider the quadratic objective of the least-cost real-time energy schedule based on Example 1. $a = [30, 0.5, 0]$ and $b = [20, 50, 1]$ represent the quadratic and primary coefficients, respectively. The lowest-cost real-time energy schedule can be formulated as:

$$\min_{P_i} \sum_{i=1}^{3} a_i P_i^2 + b_i P_i \tag{3a}$$

$$\text{s.t.} \ 0 \leq P_i \leq \bar{P}_i, \quad i = 1, 2, 3, \tag{3b}$$

$$\sum_{i=1}^{3} P_i = \hat{L}, \tag{3c}$$

**Example 3**: We consider reserve procurement based on Example 1. The lowest-cost day-ahead energy schedule can be formulated as:

$$\min_{P_i, R_i^U, R_i^D} \sum_{i=1}^{3} C_i P_i + C_i^U R_i^U + C_i^D R_i^D \tag{4a}$$

$$\text{s.t.} \ \sum_{i=1}^{3} R_i^U = s^U \hat{L}, \ \sum_{i=1}^{3} R_i^D = s^D \hat{L}, \tag{4b}$$

$$R_i^D \leq P_i \leq \bar{P}_i - R_i^U, \quad i = 1, 2, 3, \tag{4c}$$

$$\sum_{i=1}^{3} P_i = \hat{L}, \tag{4d}$$

where $R^U = [R_1^U, R_2^U, R_3^U]$ and $R^D = [R_1^D, R_2^D, R_3^D]$ represent the upward and downward reserve provision for the generator, respectively, while $C^U = [40, 40, 40]$ \$/MW and $C^D = [30, 30, 30]$ \$/MW denote the corresponding reserve cost coefficients. Reserve requirements are based on a certain percentage of the load, with the proportional coefficients of upward and downward reserves being represented as $s^U$ and $s^D$, respectively. Once reserve allocation $R^{U*}$ and $R^{D*}$ are determined, the lowest-cost real-time energy schedule can be formulated as:

$$\min_{L_{sh}, r_i^U, r_i^D} C_{VoLL} L_{sh} + \sum_{i=1}^{3} C_i (r_i^U - r_i^D) \tag{5a}$$

$$\text{s.t.} \ 0 \leq r_i^U \leq R_i^{U*}, \ 0 \leq r_i^U \leq R_i^{U*}, \quad i = 1, 2, 3, \tag{5b}$$

$$L_{sh} + \sum_{i=1}^{3} r_i^U - r_i^D \geq L - \hat{L} \tag{5c}$$

where $r^U$ and $r^D$ represent the upward and downward reserve deployment for the generator. The proportional coefficients $S^U$ and $S^D$ are set at 0.15 each. The impact of the forecast error on the total cost increment is shown in Fig. 2(d).

These three examples illustrate that using the expected value to make a decision is not necessarily the optimal choice. The fundamental reason behind this problem is that the expected value ignores the impact of parameter uncertainty on decision objectives. For instance, if the impact of forecast error on the decision objective is known prior to making the decision, such as the black line in Fig. 2(b), then slightly over-forecasting than the expected value can potentially lead to a more optimal decision under expectation since it is uncertain whether the expected value is over- or under-forecasted. These findings illustrate that the optimal decision must take into account any uncertainty in the parameters, rather than solely focusing on the most likely values. DFL provides a viable solution to tackle this problem.